\documentclass[a4paper, 12pt]{article} % Font size (can be 10pt, 11pt or 12pt) and paper size (remove a4paper for US letter paper)
\usepackage[right=1in,left=1in,top=1in,bottom=1in]{geometry}
\usepackage{amsmath}
\usepackage[utf8]{inputenc}
\usepackage[T1]{fontenc} 
\usepackage{pxfonts}
\usepackage{tikz}   
\usepackage{tikz-3dplot}
\usepackage{tcolorbox}
\usepackage{pgfplots}
\usepackage[titles]{tocloft}
\newcommand{\x}[1]{{#1}}

\title{\Large{Algorithmic Randomness and Probabilistic Laws}}
\author{Jeffrey A. Barrett\footnote{Department of Logic and Philosophy of Science, University of California, Irvine, Irvine, CA 92697-5100. Email: j.barrett@uci.edu}
~~and
Eddy Keming Chen\thanks{Department of Philosophy,  University of California, San Diego, 9500 Gilman Dr, La Jolla, CA 92093-0119. Email: eddykemingchen@ucsd.edu  }}
\date{\textit{The British Journal for the Philosophy of Science}, accepted version}

\begin{document}

\maketitle

\begin{abstract}
We apply recent ideas about complexity and randomness to the philosophy of laws and chances. We develop two ways to use algorithmic randomness to characterize probabilistic laws of nature. The first, a \emph{generative $\mbox{chance}^{\star}$ law}, employs a nonstandard notion of chance. The second, a \emph{$\mbox{probabilistic}^{\star}$ constraining law}, impose relative frequency and randomness constraints that every physically possible world must satisfy. The constraining notion removes a major obstacle to a unified governing account of non-Humean laws, on which laws govern by \emph{constraining} physical possibilities; it also provides independently motivated solutions to familiar problems for the Humean best-system account (the Big Bad Bug and the zero-fit problem). On either approach, probabilistic laws are tied more tightly to corresponding sets of possible worlds: some histories permitted by traditional probabilistic laws are now ruled out as physically impossible. Consequently, the framework avoids one variety of empirical underdetermination while bringing to light others that are typically overlooked.
\end{abstract}

\tableofcontents
%\vspace{10pt} 

\section{Introduction}

Probabilistic laws, as they are usually understood, involve a variety of underdetermination. This is illustrated by a simple example.

Consider repeated tosses of a coin that produce an infinite $\omega$-sequence of results $\langle r_1, r_2, \ldots \rangle$, where $r_i$ is the result of the $i$th coin toss. Each such possible sequence of tosses gives the history of events of a possible world. Let $\Omega^{L}$ be the set of all such worlds that accord with a law $L$.\footnote{\x{In this toy model the $\omega$-sequence is the world's entire \emph{mosaic}.  A mosaic, in Lewis’s sense, is the total distribution of (localized) categorical facts across space-time (Lewis 1986, p.~ix); because space-time structure plays no role for coin-tosses, the mosaic in our example is just the outcome sequence itself. Both Humean and non-Humean theories can represent the \emph{physical possibilities} for a candidate law \(L\) by a class \(\Omega^{L}\) of such mosaics (or ``worlds'' in the main text; cf. the definition of Laplacian determinism by Earman 1986, p.~22). The two camps diverge on \emph{supervenience}: Humeans hold that the laws supervene on the actual mosaic, whereas Minimal Primitivism (MinP) and other non-Humean views deny this (Chen and Goldstein 2022; Chen 2024b). On MinP, a complete specification of a possible world is the ordered pair \((\xi, L)\) of a mosaic \(\xi\) and its governing law \(L\); the physical possibilities of \(L\), \(\Omega^{L}\), are defined to be exactly the set of mosaics \(\xi\) that satisfy \(L\).  Other non-Humean constraint accounts---including Adlam's modal-constraint view (Adlam 2022) and Meacham's nomic-likelihood view (Meacham 2023)---use the same mosaic-based definition, which is the framework adopted here.}}

Now consider the probabilistic law $L$: 
\begin{description}
    \item[$L$:] Each element in the $\omega$-sequence of coin tosses $\langle r_1, r_2, \ldots \rangle$ is determined independently and with an unbiased probability of heads and tails.
\end{description}
One might think of $L$ as descriptive of a fundamentally random process, something like starting with a sequence of spin-$1/2$ particles each in a eigenstate of $z$-spin, then measuring their $x$-spins in turn.

As probabilistic laws are typically understood, $\Omega^{L}$ is the set of all $\omega$-sequences. That is, $L$ does not rule out any world. A world compatible with $L$ might exhibit \emph{any} limiting relative frequency or no limiting relative frequency at all. As a result, even the full \x{event} history of a world will fail to determine $L$ in an \x{uncountably infinite} cardinality of cases. And since $\Omega^{L}$ is compatible with every probabilistic law with heads and tails as possible outcomes with positive probability on each toss, even the full set of worlds compatible with $L$ does nothing to determine $L$ over any other probabilistic law.

% This sort of underdetermination is closely related to a corresponding sort of empirical coherence.\footnote{See Barrett (1996), (1999) and (2020) for a presentation and discussions of empirical coherence.} A physical law is empirically coherent, in the sense we are interested in here, only if it is always in principle possible for one to have empirical support for the law if the law is in fact true.\footnote{Throughout this paper unless specified otherwise, when we say ``$L$ is the true law'' or ``the law $L$ is true,'' we mean not just $L$ is true but also $L$ is the law. This is compatible with the non-Humean perspective where laws govern and the Humean perspective where laws form the optimal description of the mosaic. } If a law is empirically incoherent, then it may be impossible to learn that the law is true with even complete evidence. The law $L$ is empirically incoherent in this sense as there are $\omega$-sequences that might occur if $L$ is true that would provide no empirical evidence whatsoever for accepting $L$. In such worlds one would never have any empirical support for accepting the correct probabilistic law even with full evidence. Indeed, since worlds compatible with $L$ might exhibit any limiting relative frequency, there is a continuous cardinality of such worlds. 

An immediate epistemic consequence of this loose link between probabilistic laws and physical possibility is that a law may be true in worlds where there can be no empirical evidence whatsoever for its acceptance.\footnote{Throughout this paper unless specified otherwise, when we say ``$L$ is the true law'' or ``the law $L$ is true,'' we mean not just $L$ is true but also $L$ is the law. This is compatible with the non-Humean perspective where laws govern and the Humean perspective where laws form the optimal description of the mosaic. For a recent survey of both perspectives, see Chen (2024b). } If so, it would be impossible to learn that the law was true in even the ideal case where one had access to complete evidence and unbounded computational resources. The law $L$ above exhibits this strong variety of underdetermination since there are $\omega$-sequences that might occur if $L$ is true that would provide no empirical evidence for accepting $L$. (Consider, for example, the all-heads sequence, the alternating heads-tails sequence, and the sequence with one-quarter heads.) And since worlds compatible with $L$ might exhibit any limiting relative frequency, there is an uncountabe infinity of such ``maverick worlds.'' 

In contrast, the link between deterministic laws and empirical evidence is much tighter. Deterministic laws do rule out worlds, and different laws are compatible with different sets of worlds. Moreover, the complete empirical evidence presented in a world often uniquely determines a deterministic law when we focus on candidate laws that are similar in complexity, such as (N-body) Newtonian gravitation theory with different values of the gravitational constant $G$. With a different value of $G$, that theory is compatible with an entirely different set of possible worlds disjoint from the original one. There is still some underdetermination with deterministic laws, but it is the standard variety that we often address (rightly or wrongly) by appealing to theoretical virtues such as simplicity.\footnote{For some discussions regarding the standard variety of underdetermination for deterministic laws, see Russell (1913, pp.22-24), Earman (1986, p.22), and Chen (2024a, section 2). \x{Now, this type of underdetermination may not be entirely foreign to deterministic theories. For example, the nomologically} \x{contingent geometrical fact about absolute motion in Newtonian mechanics is  underdetermined by observations of relative positions. What is special with standard probabilistic theories is the strong underdetermination of the laws by complete evidence about the mosaic.  We thank an anonymous referee for comments here.}} Different yet equally complex probabilistic laws such as $L$ that vary only in their biases towards heads are compatible with the same set of possible worlds---$\Omega^{L}$.

One might get a tighter fit between probabilistic laws and empirical evidence by appealing to a stronger conception of probabilistic laws, one more closer aligned with the epistemic situation presented by deterministic laws.
%that presents no epistemic problems beyond those one already faces in learning a deterministic law. [i commented this out because there is a new sort of computational problem one faces with star laws in dsitinguishing between L*MLR and L*SR say.]
To this end, consider the law $L^{\star}$: 
\begin{description}
    \item[$L^{\star}$:] The $\omega$-sequence of coin tosses $\langle r_1, r_2, \ldots \rangle$ is \emph{random} with unbiased relative frequencies of heads and tails.
\end{description}
Here being \emph{random} is a property of the $\omega$-sequence. It remains then to say what it might mean for a sequence to be random.

The notions of randomness we will consider here are algorithmic. They are defined in terms of statistical tests that determine whether an $\omega$-sequence exhibits an \x{effectively} specifiable pattern. What matters at present is that each sequence will either pass or fail a particular test for being random.

\begin{figure}
    \centering
    \begin{tikzpicture}
\draw[black, ultra thick] (0,0) rectangle (5,4);  
\draw[blue, ultra thick] (0.2,0.2) rectangle (4.75,3.5);
\node[blue] at (0.7,2) {$\Omega^{L^{\star}}$};
\node[black] at (-0.32,2) {$\Omega^{L}$};
\node[red] at (5.6,3.8) {$M^{L}$};
\fill[fill=red!25] (0,0) rectangle (0.2,4);
\fill[fill=red!25] (0,0) rectangle (5,0.2);
\fill[fill=red!25] (4.75,0) rectangle (5,4);
\fill[fill=red!25] (0,3.5) rectangle (5,4);
\draw[->, thick] (5.3,3.7) -- (4.87,3.7);
\end{tikzpicture}
    \caption{$\Omega^{L}$, the set of worlds compatible with law $L$, is the set of all $\omega$-sequences of coin toss results. $\Omega^{L^{\star}}$, the set of worlds compatible with law $L^{\star}$, is a proper subset of $\Omega^{L}$. All members of $\Omega^{L^{\star}}$ exhibit the random pattern and relative frequencies stipulated by $L^{\star}$. $M^L$, the relative complement of $\Omega^{L^{\star}}$ in $\Omega^{L}$, is the set of `maverick worlds,' i.e. those that are usually regarded as compatible with $L$ but lack the random pattern or relative frequencies.}
    \label{fig:my_label}
\end{figure}
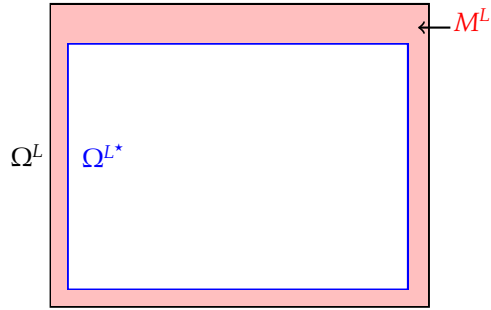

While $L$ is compatible with all $\omega$-sequences of results, $L^{\star}$ is not. Let $\Omega^{L^{\star}}$ be the set of all worlds that accord with the law $L^{\star}$. All worlds in $\Omega^{L^{\star}}$ exhibit the random unbiased sequences stipulated by $L^{\star}$ and hence, unlike $\Omega^{L}$, is a proper subset of the set of all possible $\omega$-sequences. Specifically, $\Omega^{L^{\star}}$ contains no maverick worlds where the results exhibit an \x{effectively} specifiable pattern or fail to exhibit the right relative frequencies or fail to exhibit any relative frequencies at all. (See Figure 1.)

%If $L^{\star}$ is true, then \emph{any} physically possible world fully determines $L^{\star}$. [note: I commented this out because if the sequence is MLR, the law might be L*SR or L*MLR.]

\x{If $L^{\star}$ is true, then no special probabilistic background assumptions or priors regarding what world one inhabits are required for successful inquiry.\footnote{Both the Principal Principle and Cournot's Principle are sometimes used for this purpose. See Diaconis and Skyrms (2018, 66--7) for a brief discussion of the latter. We return to this point in \S5.} An inquirer in any physically possible world might determine the truth of $L^{\star}$ in the limit of inquiry by considering the results of coin tosses. Indeed, the $\star$-laws we consider here are empirically discoverable in the strong sense that, with complete evidence, a good Bayesian inquirer who takes the law seriously to begin may gain evidence in favor of the law in the short to medium run and will surely learn it in the limit of inquiry up to an equivalence class of computationally indistinguishable laws, an epistemic degree of freedom we will discuss later.\footnote{\x{To learn $L^*_{ML}$, for example, the inquirer might update her credence that each initial segment is $c$-incompressible on a prefix-free Turing machine. Depending on the inquirer's priors, the inquirer might take herself to have evidence in favor of the law if initial sequences are $c$-incompressible for a given $c$ for a long time. Further, two computationally limited Bayesian agents with different priors that agree on which data streams are algorithmically random will agree asymptotically in their posteriors, displaying a robust merging of opinion in the Bayesian sense (Zaffora Blando, 2022).}} In this, a probabilistic law like $L^\star$ is much like a deterministic law.}

%A probabilistic law like $L^\star$ is much like a deterministic law in that the law is fully determined by the evidence.\footnote{\x{That is, complete evidence fully determines the law up to the standard variety of underdetermination discussed earlier.}} [note: I comment this out because full evidence typically does not distngush between different *star laws*. If L*ML is true, then one would expect an omega sequence that is entirely compatible with L*SR, for example.]

In the present paper, we consider how one might understand a law like $L^{\star}$ as a \emph{generative $\mbox{chance}^\star$ law} or as a \emph{$\mbox{probabilistic}^\star$ constraining law}. While the two notions are closely related, we favor thinking of a law like $L^{\star}$ as a $\mbox{probabilistic}^\star$ constraining law that governs the full sequence of coin-tosses. That said, both notions are relevant to discussions regarding the metaphysics of laws generally.

This approach represents a nonstandard way of understanding probabilistic laws that has several salient virtues. As we explain in \S4, the notion of a $\mbox{probabilistic}^\star$ constraining law removes a major obstacle for developing a unified non-Humean account of governing laws, according to which laws govern by constraining  physical possibilities. Such a notion also provides independently motivated solutions to the issues of the Big Bad Bug and the definition of \emph{fit} in the Humean best-system account of laws. Hence, the new notion of probabilistic law can be used to address long-standing issues in the metaphysics of laws and chances. The finer-grained distinctions available on this approach also reveal an underappreciated variety of empirical underdetermination. Not only are $L$ and $L^{\star}$ in a strong sense empirically equivalent, but there are  incompatible versions of $L^{\star}$ that are also empirically equivalent if one is limited to Turing-strength computations. Finally, we will discuss a sense in which laws like $L^{\star}$ exhibit a subtle variety of temporal non-locality. Though unfamiliar and counterintuitive at first, $\star$-laws are worthy of study and may lead to new ideas concerning the nature of laws.

\section{Randomness constraints}

In order to characterize a $\star$-law, one needs a test of randomness for $\omega$-sequences. A random sequence of tosses with an unbiased coin should exhibit an even relative frequency of heads and tails in the limit. But this, of course, is not sufficient. The limiting relative frequency of an alternating sequence of heads and tails will be $1/2$ for heads and tails, but this sequence is clearly not random.

There are three further conditions that one should want an unbiased random sequence to satisfy: a random sequence should be \x{patternless, typical given the limiting relative frequencies}, and not allow for the success of a fair betting strategy.\footnote{See Li and Vit\'anyi (2019) and Dasgupta (2011) for introductions to algorithmic complexity and randomness. See Barrett and Huttegger (2021) for a discussion of these notions and how they relate to each other. The present section follows part of that discussion. See Eagle (2021) for an introduction to some of the philosophical issues involving randomness, and Hajek (2023) for an introduction to the closely related topic, interpretations of probability.} These conditions are closely related. The core idea is that an $\omega$-sequence should count as random only if it exhibits no effectively specifiable regularity that characterizes the sequence and might consequently be used to make predictions better than chance.

Algorithmic tests are helpful in characterizing what it might mean for an infinite sequence to be patternless. As a first try, one might take an $\omega$-sequence to be patternless, and hence random, if and only if there is no finite-length algorithm that produces the sequence.\footnote{One might think of an algorithm as a program in a Turing-computable language and the length of the algorithm as the length of the program. Different languages will differ in the length they assign to an abstract algorithm by no more than the length of a program that translates between the two languages.} If there is such an algorithm for an $\omega$-sequence of coin-tosses, then the algorithm expresses a regularity, something that one might even think of as a deterministic law, that characterizes the sequence. But that an $\omega$-sequence cannot be represented by a finite algorithm is again not sufficient for it to be random in the sense we are interested in here.

Consider an infinite sequence that consists of a repeated three-block pattern of one thousand heads followed by one thousand tails followed by one thousand random and unbiased heads and tails. The relative frequency of heads and tails in the full sequence is unbiased. And since there are an infinite number of random blocks, such a sequence cannot be represented by a finite-length algorithm. But the sequence is clearly not random. A good Bayesian inquirer might quickly learn to bet on heads a thousand times, then bet on tails a thousand times, then bet anything at all a thousand times, then repeat the pattern. If so, she will enjoy unbounded wealth in the limit.

The problem here is that there is no bound on the amount that a finite initial segment of this sequence might be compressed. One might write a very short program that takes advantage of the regularity of the blocks of heads and tails, then write a program that outputs an initial segment by alternating that short routine with a routine that just lists each random block. In this way, one might eventually shorten the algorithmic representation of finite initial segments of the sequence by more than any constant~$c$. This observation provides the key idea behind \emph{Martin-L\"of randomness}.

An $\omega$-sequence is Martin-L\"of random if and only if there is a constant~$c$ such that all finite initial segments are $c$-incompressible by a prefix-free Turing machine.\footnote{An initial segment is $c$-incompressible if and only if it is not representable by an algorithm that is $c$~shorter than the initial segment. A prefix-free Turing machine is a universal Turing machine that is self-delimiting and hence can read its input in one direction without knowing what, if anything, comes next. Such a machine provides an even playing field. See Li and Vit\'anyi (2019).} This definition also satisfies the two other desiderata for a suitable notion of randomness. If a sequence is Martin-L\"of random, then there is no fair betting strategy that generates unbounded wealth. And since measure one of infinite-length sequences are Martin-L\"of random in unbiased Lebesgue measure, it meshes well with the intuition that random sequences are \x{typical given the limiting relative frequencies}.

One might also define what it means for a sequence to be Martin-L\"of random by considering the set of statistical tests that such a sequence will pass. A Martin-L\"of test is a sequence $\{U_n\}_{n\in \omega}$ of uniformly $\Sigma^0_1$ classes such that $\mu(U_n) \leq 2^{-n}$ for all $n$, where $\mu$ is the unbiased Lebesgue measure over the sequences. Being uniformly $\Sigma^0_1$ means that there is a single constructive specification of the sequence of classes. A constructive specification is one that can be represented by an ordinary algorithm.\footnote{See Barrett and Huttegger (2021) for further details.} The idea is that each sequence $\{U_n\}_{n\in \omega}$ of uniformly $\Sigma^0_1$ classes corresponds to a way that a sequence might be special and thus fail an associated statistical test of randomness. A sequence passes a particular Martin-L\"of test if it is not special in the specified sense.

Let $2^\omega$ be the set of all $\omega$-length sequences (infinite-length sequences indexed by ordinal $\omega$). A class $C \subset 2^\omega$ is Martin-L\"of null if there is a Martin-L\"of test$\{U_n\}_{n\in \omega}$ such that $C \subseteq \bigcap_n U_n$. A sequence $S \in 2^\omega$ is \emph{Martin-L\"of random} if and only if $\{S\}$ is not Martin-L\"of null. That is, a sequence $S$ is Martin-L\"of random if and only if it passes every Martin-L\"of test. And again, a sequence has this property if and only if there is a constant~$c$ such that all finite initial segments are $c$-incompressible by a prefix-free Turing machine.

One might use the notion of Martin-L\"of randomness to specify the law $L^{\star}$ as a constraint on the set of physically possible worlds:
\begin{description}
    \item[$L^{\star}_{ML}$:] The $\omega$-sequence of coin tosses $\langle r_1, r_2, \ldots \rangle$ is \emph{Martin-L\"of random} with unbiased relative frequencies of heads and tails.
\end{description}
Here \emph{all} of the worlds in $\Omega^{L^{\star}_{ML}}$ are random with well-defined unbiased relative frequencies.\footnote{\x{As we are using the basic notion of Martin-L\"of randomness, if the sequence is random, the relative frequencies are guaranteed to be unbiased, so that clause in the law is redundant. For biased processes, one needs a more general notion of algorithmic randomness. Martin L\"of himself showed how to provide an algorithmic notion corresponding to an arbitrary computable probability distribution (1966, 612--4).} \x{See Porter (2020) for a recent survey of work on biased algorithmic randomness.}} As a result, a non-dogmatic inquirer will surely infer unbiased relative frequencies in the limit. And inasmuch as all initial segments of her data will be $c$-incompressible, she will have as good of evidence as possible that the data are patternless and are hence randomly distributed.\footnote{A $\mbox{probabilistic}^\star$ law need not presume a fundamental direction of time. In order to determine an initial segment, the definition of Martin-L\"of randomness seems to presuppose an initial time and a temporal direction. However, one can generalize the notion by requiring that the ordered-sequence of coin tosses be Martin-L\"of random for any specified temporal direction and for any toss one regards as the ``initial'' toss.}

\section{Alternative algorithmic notions}

Martin-L\"of randomness is not the only way that one might characterize a probabilistic coin-toss law. There are other algorithmic notions of randomness to choose from. Schnorr randomness is a closely-related notion with many of the same virtues.

A Schnorr test is a Martin-L\"of test where the measures~$\mu(U_n)$ are themselves uniformly computable. A class $C \subset 2^\omega$ is Schnorr null if there is a Schnorr test $\{U_n\}_{n\in \omega}$ such that~$C \subseteq \bigcap_n U_n$. And a sequence $S \in 2^\omega$ is \emph{Schnorr random} if and only if $\{S\}$ is not Schnorr null.

\begin{figure}
    \centering
    \begin{tikzpicture}
\draw[black,  very thick] (0,0) rectangle (5,4);  
\draw[blue,  very thick] (0.3,0.3) rectangle (4.6,3.4);
\draw[purple,  very thick] (0.15,0.15) rectangle (4.8,3.7);
\node[blue] at (0.8,2) {$\Omega^{L^{\star}_{ML}}$};
\node[purple] at (5.55,2) {$\Omega^{L^{\star}_{S}}$};
\node[black] at (-0.32,2) {$\Omega^{L}$};
%\node[red] at (5.6,3.8) {$M^{L}$};
%\fill[fill=red!25] (0,0) rectangle (0.2,4);
%\fill[fill=red!25] (0,0) rectangle (5,0.2);
%\fill[fill=red!25] (4.75,0) rectangle (5,4);
%\fill[fill=red!25] (0,3.5) rectangle (5,4);
\draw[->, thick] (5.2,2) -- (4.82,2);
\end{tikzpicture}
    \caption{$L^{\star}_{S}$, which is formulated with Schnorr randomness, is compatible with more worlds than $L^{\star}_{ML}$, which employs Martin-L\"of randomness. However, the two are empirically indistinguishable, if one is limited to Turing-strength computation.  }
    \label{fig:my_label}
\end{figure}
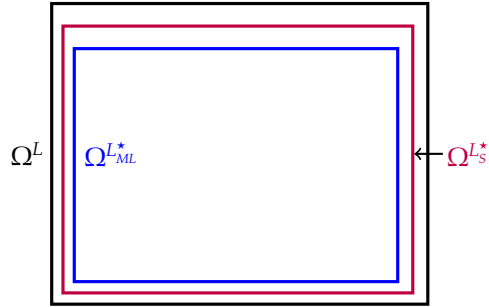

Schnorr randomness has similar virtues to Martin-L\"of randomness. Initial segments are patternless in a strong sense, there is a natural sense in which there is no fair betting strategy, and measure one of infinite-length sequences, including all those that are Martin-L\"of random, are Schnorr random. (See Figure 2.) And as with Martin-L\"of randomness, one might use the notion of Schnorr randomness to specify a probabilistic constraining law:
\begin{description}
    \item[$L^{\star}_{S}$:] The $\omega$-sequence of coin tosses $\langle r_1, r_2, \ldots \rangle$ is \emph{Schnorr random} with unbiased relative frequencies of heads and tails.
\end{description}
One might also consider an associated notion of chance here as well. A $\mbox{chance}^{\star}_{S}$ process behaves just like an ordinary chance process except that it can never produce an infinite sequence that fails to be Schnorr random with well-defined relative frequencies.

While Martin-L\"of randomness provides a particularly natural notion of randomness, Schnorr randomness also has conceptual virtues.\footnote{See Downey and Griffiths (2004) for details regarding the properties of Schnorr randomness and Downey and Hirschfeldt (2010) for a description and comparison of Martin-L\"of and Schnorr randomness.} So there is a choice to make, but it is arguably a choice without empirical consequences.

Since there are sequences that are Schnorr random but not Martin-L\"of random, $L^{\star}_{ML}$ and $L^{\star}_{S}$ are different laws. That said, they are in a strong sense empirically equivalent since there is no effective procedure that would determine whether a particular sequence is Martin-L\"of random or Schnorr random but not Martin-L\"of random.\footnote{\x{The reason is that the set of Martin-L\"of random sequences and the set of Schnorr random but not} \x{Martin-L\"of random sequences are both \textit{tail sets}. In consequence, no finite sample of the sequence will determine whether it is ML random or Schnorr random but not Martin-L\"of random. Hence, there is no effective procedure that will determine the set to which it belongs. The notion of effective procedure here, as elsewhere in the paper when not explicitly stipulated to be otherwise, is the standard Church-Turing one. See Barrett and Huttegger (2021) for a discussion.}} Hence if one is limited to Turing-strength computation, one would never be able to distinguish between $L^{\star}_{ML}$ and $L^{\star}_{S}$ no matter what empirical evidence one had. \x{This is the computational equivalence between alternative $\star$-laws mentioned earlier.}

The upshot is that moving from a standard probabilistic law to a probabilistic constraining law eliminates one variety of empirical underdetermination, but it reveals two others. First, insofar as one expects a sequence of coin tosses governed by a traditional probabilistic law $L$ to be such that one can detect no discernible pattern, one should expect \x{(either kind of)} $L^\star$ to be empirically indistinguishable from $L$. And insofar as one is limited to Turing-strength computations, one will be unable to distinguish between different \x{algorithmic} versions of $L^\star$ like $L^{\star}_{ML}$ and $L^{\star}_{S}$. Since Martin-L\"of randomness has the sort of properties we want (such as the equivalence with two other intuitively correct notions of randomness), and as it is arguably the standard algorithmic notion (Dasgupta 2011), we shall understand $L^{\star}$ as $L^{\star}_{ML}$.

\section{$\mbox{Chances}^{\star}$, $\mbox{probabilities}^{\star}$, and \x{credences}}

What kind of physical law is $L^{\star}$? And how does it govern the world? Again, one might think of $L^{\star}$ as a \emph{generative $\mbox{chance}^{\star}$ law} or as a \emph{$\mbox{probabilistic}^{\star}$ constraining law}. We will start with the first.

As a \emph{generative $\mbox{chance}^{\star}$ law}, $L^{\star}$ tells us that each toss is generated by unbiased $\mbox{chances}^{\star}$, where a $\mbox{chance}^{\star}$ process behaves just like an ordinary chance process except that it can \emph{never} produce an infinite sequence that fails to be Martin-L\"of random or fails to exhibit well-defined relative frequencies.\footnote{\x{Note that \textit{generated by unbiased $\mbox{chances}^{\star}$} in the case of $L^\star_{ML}$ means \textit{will yield a Martin-L\"of random sequence with unbiased limiting relative frequencies.} This is just what $\mbox{chances}^{\star}$ are.}}
% Alternatively, one might consider a similar algorithmic notion of chance but without requiring there to be well-defined relative frequencies. Here we are assuming well-defined relative frequencies so that an agent might infer the law given full empirical evidence by conditioning on the results of coin tosses as she goes. The cost of this further constraint is modest since we are already requiring the full sequence to be Martin-L\"of random.} [note: I commented this out as it adds an unecessary complication. It would not be too difficult to put it back in if needed.]
As a result, a $\mbox{chance}^{\star}$ process involves a subtle violation of independence. The sequence of tosses will pass every effective test for statistical independence, but since the full sequence must exhibit the property of being Martin-L\"of random with unbiased relative frequencies, a $\mbox{chance}^{\star}$ process is \emph{holistically} constrained. The constraint is not felt on any finite set of tosses, nor is it discoverable by effective means, but it does require that a relationship hold between the full sequence of tosses that is generated by the process in the limit. This \x{non-local} interdependence between outcomes may be incompatible with the usual \x{reasons for wanting} a generative law. It may also be incompatible with how causal explanation works more generally.\footnote{Nevertheless, the temporal non-locality here is more subtle and less dramatic than the kind found in certain theories in quantum foundations. Consider, for example, the retrocausal model of Sutherland (2008), where the particle velocity right now is partly determined by a wave function evolving from the future. As a generative $\mbox{chance}^{\star}$ law, $L^{\star}$ requires no physical entity evolving from the future to the present moment. Instead, we may regard the \emph{generative $\mbox{chance}^{\star}$ mechanism} as something that uses Martin-L\"of random sequences to determine the chance outcomes step by step. For a recent discussion about temporal non-locality in quantum foundations, see Adlam (2018) and the references therein.} 

Given this, $L^{\star}$ is perhaps more naturally regarded as a law that governs by constraining the entire history of the world---in this case, the full $\omega$-sequence of outcomes. It tells us which sequences of outcomes are physically possible, namely those that satisfy the specific type of randomness constraint imposed by the law. We elaborate on $\mbox{probabilistic}^{\star}$ constraining law in the next section.\footnote{Three notes about the literature. (1) The constraint account fleshes out one of the interpretive options of probabilistic laws discussed by Chen and Goldstein (2022, \S3.3.3, Option 4).  (2) In one respect, the present account is similar to John T. Roberts's nomic frequentism (2009), as they both employ frequency constraints. However, Roberts does not appeal to algorithmic randomness. On nomic frequentism, non-random sequences (such as the alternating heads-tails sequence) are still physically possible. Inasmuch as any non-random sequence is regarded as evidence against the probabilistic law and in favor of a deterministic law, it would be better to exclude such sequences from physical possibilities. (3)  Adlam (2022) presents an account of laws of nature as constraints that is similar to Chen and Goldstein (2022) and contains a helpful discussion of Roberts's nomic frequentism.}

\x{Neither way of understanding $\star$-laws automatically delivers single-shot credences. That said, depending on the symmetries exhibited by her priors, an agent who is committed to a $\star$-law may find that she is also committed to single-shot credences that accord with the limiting relative frequencies associated with the law.

Suppose that an agent believes $L^\star_{ML}$ for a sequence of fair coin tosses. Regardless of whether she understands it as a generative $\mbox{chance}^{\star}$ law or as a $\mbox{probabilistic}^{\star}$ constraining law, a commitment to the law does not, by itself, determine her credences for the result of the next coin toss. Indeed, she might take the full sequence of tosses to be Martin-L\"of random and at the same time consistently believe that the next ten tosses will be heads. This is because it is the \emph{tail} of the sequence, not any finite initial segment, that makes a sequence Martin-L\"of random. But if the agent believes $L^\star_{ML}$ and if her conditional credences are \emph{exchangeable} (that is, if she, conditional on $L^\star_{ML}$,  assigns the same credence to each initial segment of a specified length independent of the order of the outcomes), then it follows from de Finetti's representation theorem that she is committed to assigning credences to each toss as if it were the result of an IID process with, in the case of $L^\star_{ML}$, unbiased classical probabilities.}\footnote{\x{See Barrett and Chen (2025) 
for an extended discussion of this point.}}

\section{\x{$\mbox{Probabilistic}^{\star}$ constraining law}}

\x{The understanding of probabilistic $\star$-laws as governing by constraining} meshes well with Chen and Goldstein's (2022) minimal primitivism account (MinP), according to which laws are certain primitive facts that govern the world by constraining the physical possibilities of the entire spacetime and its contents.\footnote{\x{A natural and general question regarding the constraint notion of laws is this: how does a law govern, if not by generating or producing outcomes? While a detailed defense of constraint-based accounts is beyond the scope of this paper and can be found in the recent literature (e.g., Chen and Goldstein 2022, Adlam 2022, Chen 2024, Meacham 2024), we offer some remarks. First, one might prefer `constraint' over `generation,' `production,' or `temporal production' based on its generality and consonance with how many fundamental laws in modern physics are understood. For example, the Einstein field equations in general relativity act as a constraint on the entire spacetime geometry, rather than temporally producing the universe step-by-step, and even permit scenarios like closed timelike curves incompatible with simple temporal evolution. Second, the challenge of explaining how a law ensures the corresponding regularity obtains in the universe---what van Fraassen terms the ``inference problem''---is not unique to constraint-based accounts. This issue arises for any non-Humean theory that posits a governing role for laws. `Production' or `necessitation' accounts (e.g., Maudlin (2007), Armstrong (1983)) also rely on primitive relations (like `brings about' or `necessitates') whose connection to the observed regularities must be stipulated. As Schaffer (2016) suggests, the question ``how does the law govern?'' prompts a choice of primitive axioms of the theory. The inference problem is a general feature of non-Humean frameworks, not a special burden for constraint accounts alone. Thanks to an anonymous referee for discussions here. }}

Understood this way, $L^{\star}$ addresses problems encountered by both non-Humean and Humean accounts of laws. We will start with the former.

On non-Humean governing accounts of laws, there is a puzzle concerning precisely how probabilistic laws govern. According to the standard view, probabilistic laws do not rule out any world. Instead, a probabilistic law such as $L$ merely assigns some number between zero and one to every (measurable) subsets in the space of all $\omega$-sequences. This raises a puzzle: what do these numbers between zero and one represent in physical reality?

Some non-Humeans appeal to gradable notions such as "propensities" (Maudlin 2007, p.20) or "probabilities of necessitation" (Armstrong 1983, p.172). Suppose a probabilistic law assigns a $0.2$ probability to the next outcome being heads. On the propensities view, the chance setup has a $0.2$ propensity to bring about a heads-outcome in the next toss. On the probabilities of necessitation view, the current state of affairs necessitates the state of affairs of a heads-outcome to $0.2$ probability. But while one might make sense of non-gradable notions of physical possibility and impossibility, gradable notions such as propensities and degrees of necessitation are less clear. This seems undesirable.\footnote{See Chen and Goldstein (2022, sections 2 and 3.3.3) for discussions.} 

In contrast, probabilistic laws, such as $L^{\star}$, can be viewed as a special class of constraining laws. They constrain what is physically possible by ruling out certain sequences of outcomes, namely the maverick worlds. A sequence is physically impossible just in case it fails either the frequency constraint or the randomness constraint imposed by the law. This allows us to do away with gradable notions such as propensities or probabilities of necessitation altogether. In their place, we require only non-gradable notions of physical possibilities and impossibilities.

Consider MinP as a non-Humean example. We can now employ a single primitive relation, namely \emph{constraining}, to understand how both probabilistic laws and non-probabilistic laws relate to the world. Both types of laws govern by constraining what is physically possible, thereby ruling out what is physically impossible. The way that $L^{\star}$ constrains the world is not so different from that of $F=ma$. $L^{\star}$ constrains the physical possibilities to be all and only the non-maverick worlds. $F=ma$ constrains the physical possibilities to be all and only the solutions of $F=ma$. In this way, $L^{\star}$ removes a major obstacle to a unified understanding of probabilistic and non-probabilistic laws.

Humeans may also find it useful to adopt $L^{\star}$ for the sort of work we have been discussing. First, it is relevant to the issue of the Big Bad Bug (Lewis 1986, pp.xiv-xvi). Lewis notices that the original version of the Principal Principle and Humean supervenience lead to a contradiction. There are certain histories of the Humean mosaic, called \emph{undermining histories}, that are assigned, according to the (Old) Principal Principle, non-zero probability, conditionalized on some probabilistic theory $T$ being the best system. However, they are also assigned, according to Humean supervenience, zero probability, because $T$ would not be the best system had any of its undermining histories been actual. 

Now, consider the sort of history that would count as undermining. An undermining history either has the wrong limiting frequencies or no limiting frequencies or exhibit patterns that can be summarized by a simpler system, such as a deterministic law in the case of the alternating heads-tails sequence. Undermining histories, then, are exactly the histories of maverick worlds, as they lack the frequency or randomness patterns exemplified by typical sequences of the standard probabilistic law.\footnote{\x{While the BSA criterion of simplicity has many dimensions, here we focus on algorithmic simplicity (and its dual, algorithmic randomness) as a principled and precise framework for evaluating probabilistic laws and their corresponding sets of models. These laws make claims about patterns in outcome sequences, and algorithmic compressibility provides a formal measure; a highly compressible mosaic history can undermine a probabilistic law's claim to be the best description. For our later arguments, we only need the claim that undermining histories are contained in the set of maverick worlds.}}

Understanding probabilistic laws as $\star$-laws rules out maverick worlds as physically impossible. If maverick worlds are physically impossible, then there are no physically possible undermining histories that can be used to derive the contradiction, and the Big Bad Bug is eliminated.\footnote{Lewis (1994, p.487) also observes:
    ``The trouble with [the Old Principal Principle] is that on the left-hand side we do conditionalise on $T$, but on the right-hand side we don't. No harm would come of this discrepancy if there weren't any undermining futures to worry about.''} 
Inasmuch as restricting to $\star$-laws is also motivated by considerations of underdetermination, a Humean may find this solution particularly natural.

While $L^{\star}$ is naturally interpreted as a constraining law that is well-suited for non-Humean governing accounts such as MinP, Humeans need not interpret the constraint as something that exists over and above the mosaic. They are free to translate the present account into their preferred language by regarding $L^{\star}$ as a new type of Humean best-system law. They might then use it to define a new notion of Humean physical possibilities ($\Omega^{L^{\star}}_{Humean}$) and regard both as supervenient on the Humean mosaic.\footnote{Hoefer (2019, pp.156-158) suggests that, in his preferred solution to the Big Bad Bug, we should conditionalize on the non-occurrence of an undermining history. Hoefer's strategy builds on (and arguably generalizes) the influential solution that replaces the (Old) Principal Principle with the New Principle that conditionalizes credence on the complete theory of chance, proposed independently by Hall (1994), Lewis (1994), and Thau (1994). Thinking of the probabilistic law as $L^{\star}$ provides a principled reason for such conditionalizations, namely that the undermining histories are physically impossible. Moreover, if a Humean adopts $L^{\star}$ as an axiom in their best system, there is no Big Bad Bug, and the New Principle reduces to the (Old) Principal Principle.} 

Second, Humeans who understand probabilistic laws as $\star$-laws can also avoid appealing to \emph{fit} as a criterion in the best-system analysis of probabilistic laws, which allows them to bypass difficulties with how to characterize this notion.\footnote{For helpful discussions, see Elga (2004).} Given an $\omega$-sequence, there is much underdetermination among probabilistic laws such as $L$, and one needs something like \emph{fit} to choose the winning best system. This is because the standard way of understanding \emph{informativeness} as the quantity of worlds being excluded does not distinguish among probabilistic laws like $L$. In contrast, there is significantly less underdetermination among probabilistic laws like $L^{\star}$. If we consider a spectrum of different probabilistic statements like $L^{\star}$ that differ, say, in their specifications of the relative frequencies in the $\omega$-sequence, then at most one of them is compatible with the $\omega$-sequence, and thus at most one of them is an axiom in the best system of that $\omega$-sequence. The best system analysis of a probabilistic law, such as $L^{\star}$, is much like that of a non-probabilistic law, such as $F=G\frac{m_1m_2}{r^2}$.  If we consider a spectrum of different versions of the Newtonian gravitational law that differ in the value of the gravitational constant $G$, then at most one of them is true of the mosaic, and thus at most one of them is an axiom of the best system of the mosaic. Given any Humean mosaic, one needs criteria such as simplicity and informativeness, but one does not need the statistical criterion of fit, to determine the best system (if there is one). Hence, the usual problems associated with fit would not arise for such Humeans.\footnote{One might wonder that there is now an analogous problem of underdetermination associated with the choice between, say, $L^{\star}_{ML}$ and $L^{\star}_{S}$. Given this, do Humeans still need something like \emph{fit}? It is not exactly analogous. The choice between them is not a problem for an idealized observer with the full sequence and sufficient computational power.}

%How single-shot credences work depends on the type of algorithmic law an agent believes. A generative $\mbox{chance}^{\star}$ law sequentially selects in an unbiased way a sequence from the set of sequences that are algorithmically random in the specified sense. If the agent believes a law like this, then her credences for the next event will match those she would have on the corresponding traditional chance law because she believes that the next event is generated in an analogous way. In contrast, a full specification of a $\mbox{probabilistic}^{\star}$ constraining law tells the agent to set her priors as if nature had selected in an unbiased way a sequence that is algorithmically random in the specified sense. If she does, her credences for the next event will match the relative frequencies exhibited by the specified type of sequence.

\section{\x{Historical connections}}

\x{Von Mises (1919) defined probability via a
\emph{collective}---an infinite binary sequence that  realizes the required
limiting frequency and remains stable under every \emph{admissible}
place–selection rule. Wald (1937) proved collectives exist once the selection rules are made \emph{countable}, and Church (1940) sharpened ``admissible'' condition to
\emph{computable}, yielding the Mises-Wald-Church account of randomness.
Unfortunately, Ville (1939) produced a sequence that passes every computable
place–selection test yet violates the law of the iterated logarithm, showing
that selection rules alone are too weak.

Martin-L\"of (1966) replaced the frequentist place–selection condition by the stronger requirement that a random sequence avoid every uniformly computably enumerable null set.  This yields a universal test,  excludes Ville’s sequence, and agrees with Kolmogorov-Chaitin incompressibility definition of randomness (by Schnorr's theorem) and the defeat of all computably enumerable martingales.  Formally, every ML-random sequence is a Mises–Wald–Church collective, but the converse fails.  It is worth emphasizing that Martin-Löf's definition is not strictly frequentist, since it explicitly depends on a background measure $\mu$. The measure is not defined in terms of limiting frequencies but a theoretical ingredient by which frequencies are judged to be effectively typical (or atypical). Thus, there is a genuine conceptual shift from the Mises–Wald–Church account to Martin-Löf randomness (and similarly to Kolmogorov-Chaitin randomness). For an accessible, philosophically informed discussion of those developments, see Dasgupta (2011).

Hájek (1997, 2009) raises several insightful objections against frequentist approaches to chances inspired by von Mises's work, two of which are potentially relevant to $L^\star$ laws, even though the $L^\star$ framework is not strictly frequentist: the reference class problem, and non-IID processes.
Using Martin-L\"of randomness, \(L^{\star}\)-program can sidestep both.
First, it evaluates the entire mosaic (a candidate physically possible world described in terms of fundamental natural properties) against all computable
tests relative to the law’s own measure, so there is no latitude to tailor a
reference class.
Second, independence plays no essential definitional role: any computable measure, including those generated by Markov processes, supports a universal test (Hoyrup and Rojas 2009, Porter 2020). Hence, non-IID $L$ laws can also have corresponding $L^\star$ laws. One can also generalize Martin-L\"of randomenss to any space of continuous paths, with uncountably many events in a single path, insofar as there is a computable measure on the space (Hoyrup and Rojas 2009).

Eagle (2004) and also  Hájek (2009) raise the issue of counterfactual independence. The idea of treating a law as a disposition, tendency, or propensity to generate a particular long-run pattern goes back to Popper’s propensity account (1959).   Jeffrey (1977), Hájek (1997, p.73) and Eagle (2004, pp.396–99) object that such long-run propensity accounts, insofar as they lay down constraints on frequencies, seem to violate \emph{counterfactual independence}. In our context, we interpret it as the worry that, once the long-run randomness constraint is fixed, outcomes in one finite block of trials can become counterfactually dependent on outcomes in a disjoint block. In a \(L^{\star}\)-law framework, this worry is sharply attenuated. Because Martin-L\"of randomness constrains only the \emph{tail behavior} of  \(\omega\)-sequences, any two finite segments remain counterfactually independent. Dependence shows up only for certain infinite sub-sequences, and only via non-effective statistical tests.  Agents limited to Turing–strength procedures therefore cannot discover or exploit the dependence. (We return to this point in \S7.)  Moreover, as discussed earlier, if an agent's priors are \emph{exchangeable} conditional on the unbiased version of \(L^{\star}\), de Finetti’s representation theorem yields single-case credences that coincide with the IID measure; \emph{a fortiori} the standard independence constraints (outcomes in one finite block of trials are probabilistically independent on outcomes in a disjoint finite block) can be satisfied at the level of credences for someone who accepts $L^\star$ laws.\footnote{See Barrett and Chen (2025).} Thus the classic independence objection (regarding finite blocks) that troubles long-run propensity theories does not undermine the \(L^{\star}\) proposal.

What has been missing in the philosophical literature is a systematic application of this mature theory of algorithmic randomness to the concept of probabilistic laws.\footnote{\x{Thanks to an anonymous referee for this framing.}} By treating \(L^{\star}\) laws as global constraints that confine physical possibility to the class of Martin–L\"of random histories, we give a precise account of probabilistic laws that explain in a similar way as deterministic laws do. In earlier sections, we have shown how this move dissolves worries about strong underdetermination, probabilistic governing, the Big Bad Bug, and the zero‐fit problem, which bridge the gap between technical work on randomness and philosophical debates over laws of nature and the foundations of probability.
}

\section{Discussion}

We have shown how a $\star$-law may be thought of as either a generative $\mbox{chance}^\star$ law or a $\mbox{probabilistic}^\star$ constraining law, where the notions of $\mbox{chance}^\star$ and $\mbox{probability}^\star$ are subtly different from traditional chance or probability. Two differences are particularly salient.

The first concerns independence. The results of coin tosses on $L^\star$ satisfy every computable test for independence and will hence appear to be statistically independent. One might say that the results are $\mbox{probabilistically}^\star$ independent. \x{Moreover, given conditionally exchangeable credences, one's subjective probabilities given $L^\star$ law will be fully independent, in the standard sense, because the conditional credence is provably an IID measure.} But inasmuch as some sequences are impossible, there is also a sense in which the results of tosses in the full $\omega$-sequence are interdependent. To understand $L^{\star}$ as a generative $\mbox{chance}^\star$ law, one would need to allow for a holistic causal structure that guarantees random sequences with unbiased relative frequencies in the limit. Depending on one's commitments regarding causal explanation, this may lead one to favor understanding $L^{\star}$ as a $\mbox{probabilistic}^\star$ constraining law. If one does decide to gives up on a generative $\mbox{chance}^\star$ law, one is, as we have just seen, left with a useful option for both proponents of governing-law accounts and Humeans.

The second difference is that $\mbox{chance}^\star$ and $\mbox{probability}^\star$ depend on a choice of a particular standard of algorithmic randomness. We saw this in the distinction between Martin-L\"of and Schnorr randomness. But if one is limited to Turing-strength computations, there is no way to distinguish between $L^{\star}_{ML}$ and $L^{\star}_{S}$ on the basis of empirical evidence alone. The result is a computational sort of empirical underdetermination.\footnote{See Barrett and Huttegger (2022) for a discussion of this point.} Since one can only learn a $\star$-law up to an equivalence class of computationally indistinguishable laws, one might take the law $L^\star$ to be any law in this class.

The possibility of $\star$-laws reveals a further variety of underdetermination. Inasmuch as one expects the sequence of tosses one gets on a traditional probabilistic law $L$ to be patternless, one expects $L$ to be computationally indistinguishable from both $L^{\star}_{ML}$ and $L^{\star}_{S}$. Of course, these varieties of empirical underdetermination are not new, they have just gone unnoticed.

That said, $\star$-laws also help to eliminate other forms of empirical underdetermination. If $L^{\star}$ is true as either a generative or constraining law, then if it is among the laws that one takes seriously, then, unlike traditional probabilistic laws but very much like deterministic laws like $F=ma$ and $F=G\frac{m_1m_2}{r^2}$, one will surely learn it on complete evidence in \emph{every} physically possible world.

In contrast, if $L$ is the true law, there will be an \x{uncountably infinite} cardinality of maverick worlds such that, if one were to inhabit any of them, one could never learn $L$ from the results of the coin tosses. On the usual approach to thinking about laws, one needs special background assumptions to overcome this difficulty. Specifically, one needs to argue that inhabiting a maverick world of the true law is sufficiently unlikely or atypical that one has rational justification for simply ignoring the possibility. One needs to know that the world one inhabits, and hence the statistical nature of the sequence of records that one has in fact recorded, is typical given the true law.\footnote{The thought is that if an agent is simply presented with a sequence of coin-toss records and knows nothing more than that she inhabits a physically possible world, then she cannot infer that the sequence exhibits typical statistical properties given the true law. Such an inference would require her to appeal to a background assumption like the Principal Principle or Cournot's Principle. In contrast, if she knows that her world is governed by a traditional generative law, she should expect that when she tosses her coin, the resulting sequence of records will exhibit typical statistical properties given that law. We leave a full discussion of this point to a future paper. %The first inferential context might be thought of as \emph{modal} and the second as \emph{forward looking}.
} While such assumptions may be warranted given one's other commitments, they are not required for finding the true law if one restricts one's hypotheses to $\star$-laws.

%For example, if we are given an actual $\omega$-sequence with unbiased relative frequencies for heads and tails, this assumption allows us to assign zero probability the possibility that according to the actual law each element in the $\omega$-sequence is determined with a $0.51$ probability of heads and a $0.49$ probability of tails, since this, and a continuous cardinality of other such laws, would regard the actual $\omega$-sequence as a maverick world.

\x{A distinctive feature of the $L^\star$ law is that it invokes a countable collection of statistical tests in defining physical possibilities. But why should the laws of nature governing the universe be bound by a condition that appeals to this sort of countable collection?\footnote{\x{Thanks to an anonymous referee for this consideration.}} This foundational question merits careful consideration. Here we offer some brief remarks. 

Algorithmic randomness notions, such as Martin-L\"{o}f randomness, represent our best formulations of what one might mean by a \textit{patternless} sequence. A probabilistic law \(L^{\star}\) that ensures randomness should exclude worlds exhibiting special patterns, and Martin-L\"{o}f randomness identifies these patterns by means of computably enumerable null tests. The sort of countable collection we consider, then, appears as a feature of the best formal tool we have for saying what one might mean by a patternless sequence. Put another way, the present project employs the naturalistic methodology of appealing to the best understood notions of patternlessness to explicate the notion of a probabilistic law. Should an alternative account of what it means for a sequence to be patternless emerge that does not appeal to the sort of collections we consider here, it might allow for an alterative, but related, account of probabilistic laws.}

In summary, $\star$-laws provide a way of understanding probabilistic laws as constraints on possible worlds. This helps to clarify how probabilistic laws might govern. For non-Humeans, they provide a unified way of thinking about laws as governing by constraints. And for Humeans, they provide a principled way that they might ignore undermining histories. Specifically, the new notion of probabilistic law provides a better and independently motivated way to deal with the Big Bad Bug.

\section{Conclusion}

We use algorithmic randomness to characterize two \x{new} types of probabilistic law: \emph{generative $\mbox{chance}^{\star}$ laws}, and \emph{$\mbox{probabilistic}^{\star}$ constraining laws}. 
We argue that $\star$-laws offer a novel way of understanding probability and chance. They help with one form of empirical underdetermination while bringing to light others that have been overlooked. For all we know, the world may be characterized by a traditional probabilistic law or one of various $\star$-laws. 

The notion of a $\mbox{probabilistic}^{\star}$ constraining law is especially attractive. First, it meshes well with the holistic character of the randomness and relative frequency constraints. Second, it addresses long-standing problems in the metaphysics of laws and chances: it directly supports a unified governing account of non-Humean laws and provides independently motivated solutions to issues in the Humean best-system account. These payoffs warrant further work at the interface of algorithmic randomness and the philosophy of science.

\section*{Acknowledgement}
We are grateful for helpful discussions with Emily Adlam, Gordon Belot, Weixin Cai, Craig Callender, Nancy Cartwright, Eugene Chua, David Danks, Kenny Easwaran, Sheldon Goldstein, Alan H\'ajek, Simon Huttegger, Kevin Kelly, Barry Loewer, Krzysztof Mierzewski, Daniel Rubio, Simon Saunders, Charles Sebens, Shelly Yiran Shi,  Teddy Seidenfeld, Brian Skyrms, Jason Turner, Isaac Wilhelm, Francesca Zaffora Blando, Nino Zangh\`i, Shimin Zhao, the editors and reviewers for the \textit{BJPS}, and audiences at the 2024 Philosophy of Science Association Biennial Meeting, the CMU / Pittsburgh Formal Epistemology Workshop, Laws of Nature Online Lecture Series, the Foundations of Probability Seminar in New York, Southern California Philosophy of Physics Group, Peking University, Lingnan University, UBC Okanagan, and UC San Diego Philosophy of Physics Reading Group. 

\vspace{1cm}

\begin{center}
\large{Bibliography}
\end{center}

\noindent
Adlam, Emily. (2018) ``Spooky Action at a Temporal Distance,'' \emph{Entropy}, \textbf{20(1)}, 41. https://doi.org/10.3390/e20010041

\vspace{.15cm}
\noindent
Adlam, Emily. (2022) ``Laws of Nature as Constraints,'' \emph{Foundations of Physics}, \textbf{52}, 28. https://arxiv.org/abs/2109.13836

\vspace{.15cm}
\noindent
Armstrong, David M. (1983) \emph{What is a Law of Nature?} Cambridge: Cambridge University Press.

\vspace{.15cm}
\noindent
Barrett, Jeffrey A. (2020) \emph{The Conceptual Foundations of Quantum Mechanics}, Oxford: Oxford University Press.

\vspace{.15cm}
\noindent
Barrett, Jeffrey A. (1999) \emph{The Quantum Mechanics of Minds and Worlds}, Oxford: Oxford University Press.

\vspace{.15cm}
\noindent
Barrett, Jeffrey A. (1996) ``Empirical Adequacy and the Availability of Reliable Records in Quantum Mechanics,'' \emph{Philosophy of Science} 63(1): 49--64.

\vspace{.15cm}
\noindent
\x{Barrett and Chen (2025) ``Algorithmic Randomness, Exchangeability, and the Principal Principle,'' manuscript.}

\vspace{.15cm}
\noindent
Barrett, Jeffrey A. and Simon Huttegger (2021) ``Quantum Randomness and Underdetermination,'' \emph{Philosophy of Science} 87(3). https://doi.org/10.1086/708712

\vspace{.15cm}
\noindent
Chen, Eddy Keming (2024a) ``Strong Determinism,'' \textit{The Philosopher's Imprint}. 

\noindent
https://doi.org/10.3998/phimp.3250

\vspace{.15cm}
\noindent
Chen, Eddy Keming (2024b)  \textit{Laws of Physics}. Cambridge University Press. 

\noindent
https://arxiv.org/abs/2309.03484

\vspace{.15cm}
\noindent
Chen, Eddy Keming and Sheldon Goldstein (2022) ``Governing without a Fundamental Direction of Time: Minimal Primitivism about Laws of Nature,'' in Yemima Ben-Menahem (ed.), \emph{Rethinking the Concept of Law of Nature}, Springer, pp.21-64. 

\noindent
https://arxiv.org/abs/2109.09226

\vspace{.15cm}
\noindent
\x{Church, Alonzo. (1940) ``On the Concept of a Random Sequence,'' \emph{Bulletin of the American Mathematical Society}, \textbf{46}(2), 130–135.}

\vspace{.15cm}
\noindent
Dasgupta, Abhijit (2011) ``Mathematical Foundations of Randomness,'' in Prasanta Bandyopadhyay and Malcolm Forster (eds.), \emph{Philosophy of Statistics (Handbook of the Philosophy of Science: Volume 7)}, Amsterdam: Elsevier, pp. 641–710. 

\noindent
http://dasgupab.faculty.udmercy.edu/Dasgupta-JSfinal.pdf

\vspace{.15cm}
\noindent
Diaconis, Persi and Brian Skyrms (2018) \emph{Ten Great Ideas about Chance}, Princeton and Oxford: Princeton University Press.

\vspace{.15cm}
\noindent
Downey, Rodney G., and Evan J. Griffiths (2004) ``Schnorr Randomness,'' \emph{The Journal of Symbolic Logic}, 69(2), pp.533-554. 

\vspace{.15cm}
\noindent
Downey, Rodney G., and Denis R. Hirschfeldt (2010) \emph{Algorithmic Randomness and Complexity}, New York, NY: Springer. 

\vspace{.15cm}
\noindent
\x{Eagle, Antony. (2004) ``Twenty‐One Arguments against Propensity Analyses of Probability,'' \emph{Erkenntnis}, \textbf{60}, 371–416. }

\vspace{.15cm}
\noindent
Eagle, Antony, (2021) ``Chance versus Randomness,'' The Stanford Encyclopedia of Philosophy (Spring 2021 Edition), Edward N. Zalta (ed.),

\noindent
https://plato.stanford.edu/archives/spr2021/entries/chance-randomness

\vspace{.15cm}
\noindent
Earman, J. (1986) \emph{A Primer on Determinism}, Volume 32. D. Reidel Publishing Company.

\vspace{.15cm}
\noindent
Elga, Adam (2004) ``Infinitesimal Chances and the Laws of Nature,'' \emph{Australasian Journal of Philosophy}, 82(1), pp.67-76. 

\vspace{.15cm}
\noindent
Hall, Ned (1994) ``Correcting the Guide to Objective Chance,'' \emph{Mind} 103: 504-517.

\vspace{.15cm}
\noindent
H\'ajek, Alan (2023) ``Interpretations of Probability,'' in Edward N. Zalta \& Uri Nodelman (eds.) \emph{The Stanford Encyclopedia of Philosophy (Winter 2023 Edition)}. 

\noindent
https://plato.stanford.edu/archives/win2023/entries/probability-interpret/

\vspace{.15cm}
\noindent
Hoefer, Carl (2019) \emph{Chance in the World: A Humean Guide to Objective Chance}, New York: Oxford University Press. 

\vspace{.15cm}
\noindent
\x{Hoyrup, Mathieu and Rojas, Crist\'obal. (2009) ``Computability of Probability Measures and Martin‐Löf Randomness over Metric Spaces,'' \emph{Information and Computation}, \textbf{207}(7), 830–847. }

\vspace{.15cm}
\noindent
\x{Jeffrey, Richard C. (1977) ``Mises Redux,'' in Robert E. Butts and Jaakko Hintikka (eds.), \emph{Basic Problems in Methodology and Linguistics}, Dordrecht: D. Reidel, 213–222.}

\vspace{.15cm}
\noindent
Lewis, David (1986) \emph{Philosophical Papers, Volume II}, Oxford: Oxford University Press.

\vspace{.15cm}
\noindent
Lewis, David (1994) ``Humean Supervenience Debugged'', \emph{Mind} 103:473-490

\vspace{.15cm}
\noindent
Li, Ming and and Paul Vit\'anyi (2019). \emph{An Introduction to Kolmogorov Complexity and Its Applications, 4th Edition}, New York, NY: Springer. 

\x{
\vspace{.15cm}
\noindent
Martin-L\"of, Per (1966) ``The definition of random sequences,'' \emph{Information and Control} 9:602--619.}

\vspace{.15cm}
\noindent
Maudlin, Tim (2007) \emph{The Metaphysics Within Physics}, New York: Oxford University Press. 

\x{
\vspace{.15cm}
\noindent
Meacham, Christopher J. (2023) ``The Nomic Likelihood Account of Laws,'' \emph{Ergo: an Open Access Journal of Philosophy} 9: 9.
}

\vspace{.15cm}
\noindent
\x{Meacham, C. J. G. (2025) ``Constraint Accounts of Laws,'' \emph{Ergo: An Open Access Journal of Philosophy}, \textbf{12}(0). https://doi.org/10.3998/ergo.7427}

\vspace{.15cm}
\noindent
\x{Mises, Richard von. (1919) ``Grundlagen der Wahrscheinlichkeitsrechnung,'' \emph{Mathematische Zeitschrift}, \textbf{5}(1–2), 52–99.}

\x{
\vspace{.15cm}
\noindent
Popper, Karl. (1959) ``A Propensity Interpretation of Probability,'' \emph{British Journal for the Philosophy of Science} 10, 25-42.
}

\x{
\vspace{.15cm}
\noindent
Porter, Christopher P. (2020) ``Biased Algoritmic Randomness'' in \emph{Algorithmic Randomness: Progress and Prospects},
Edited by Johanna N.\ Y.\ Franklin and Christopher P.\ Porter. Lecture Notes in Logic, 50, Association for Symbolic Logic.}

\vspace{.15cm}
\noindent
Roberts, John T. (2009) ``Laws about Frequencies,'' preprint, 
http://philsci-archive.pitt.edu/505

\vspace{.15cm}
\noindent
Russell, B. (1913) ``On the notion of cause.'' \emph{Proceedings of the Aristotelian society}, 13:1–26.

\vspace{.15cm}
\noindent
\x{Schaffer, Jonathan. (2016) ``It Is the Business of Laws to Govern,'' \emph{Dialectica}, \textbf{70}(4), 577–588.}

\vspace{.15cm}
\noindent
Sutherland, Roderich Ian (2008) ``Causally Symmetric Bohm Model,'' \emph{Studies in History and Philosophy of Science Part B: Studies in History and Philosophy of Modern Physics,} 39(4): 782-805 

\vspace{.15cm}
\noindent
Thau, Michael (1994) ``Undermining and Admissibility'', \emph{Mind} 103: 491-503

\vspace{.15cm}
\noindent
\x{Ville, Jean. (1939) \emph{\'Etude critique de la notion de collectif}. Paris: Gauthier-Villars.}

\vspace{.15cm}
\noindent
\x{Wald, Abraham. (1937) ``Zur Theorie der Kollektive,'' \emph{Mathematische Zeitschrift}, \textbf{41}, 77–92.}

\vspace{.15cm}
\noindent
\x{Zaffora Blando, Francesca. (2022) ``Bayesian Merging of Opinions and Algorithmic Randomness,'' \emph{The British Journal for the Philosophy of Science}. https://doi.org/10.1086/721758}

\end{document}